\documentclass[%
 reprint,
superscriptaddress,
 amsmath,amssymb,
 aps,
prb]{revtex4-2}

\usepackage{graphicx}% Include figure files

\usepackage{epstopdf}
\AppendGraphicsExtensions{.tif}

\usepackage{dcolumn}% Align table columns on decimal point
\usepackage{bm}% bold math
\usepackage{amsmath}
\usepackage{tikz}
\usetikzlibrary{shapes.geometric, arrows}
\usepackage{chemfig}
\usepackage{braket}
\usepackage{nicefrac}
\usepackage{bm}% bold math
\usepackage{amsmath}
\usepackage{amsmath}
\usepackage{multirow}
\usepackage{booktabs}
\usepackage{epsfig}
\usepackage{xcolor,soul}
\usepackage{epstopdf}
\usepackage{lipsum,adjustbox}
\usepackage{breakcites}
\begin{document}
\preprint{APS/123-QED}

\title{Fast calculation of retarded potentials in multi-domain TDDFT}

\author{Matan Shapira}
\affiliation{Department of Physical Electronics, Tel Aviv University, Tel Aviv 69978, Israel}
\email{matanshapira@mail.tau.ac.il}

\author{Vitaliy Lomakin}
\affiliation{Department of Electrical and Computer Engineering, University of California, San Diego, La Jolla, CA 92093, USA}
\author{Amir Boag}
\affiliation{Department of Physical Electronics, Tel Aviv University, Tel Aviv 69978, Israel}
\email{boag@eng.tau.ac.il}

\author{Amir Natan}
\affiliation{Department of Physical Electronics, Tel Aviv University, Tel Aviv 69978, Israel}
\affiliation{The Sackler Center for Computational Molecular and Materials Science, Tel Aviv University, Tel Aviv 69978, Israel}
\email{amirnatan@post.tau.ac.il}

\date{\today}

\begin{abstract}
A formulation for the efficient calculation of the electromagnetic retarded potential generated by time-dependent electron density in the context of real-time time dependent density functional theory (RT-TDDFT) is presented. The electron density is considered to be spatially separable, which is suitable for systems that include several molecules or nano-particles. The formulation is based on splitting the domain of interest into sub-domains and calculating the time dependent retarded potentials from each sub-domain separately. The computations are accelerated by using the fast Fourier transform and parallelization. We demonstrate this formulation by solving the orbitals dynamics in systems of two molecules at varied distances. We first show that
for small distances we get exactly the results that are expected from non-retarded potentials, we then show that for
large distances between sub-domains we observe substantial retardation effects.
\end{abstract}
%\end{frontmatter}}]
\maketitle

\section{Introduction}
Understanding the response of large molecular systems to an external electromagnetic fields is critical for the design and analysis of many nano-scale devices, such as quantum dots, nano-ribbons, and nano-antennas \cite{donchev2014rectenna,slepyan2017quantum,yamada2018time}. Density functional theory (DFT) \cite{koch2015chemist,sham1966one,hohenberg1964inhomogeneous} and time dependent density functional theory (TDDFT) \cite{runge1984density,casida2012progress,ullrich2012time} are broadly used methods for the calculation of ground state and excited state properties of molecules, clusters, and solid materials. The DFT Kohn-Sham (KS) equations \cite{kohn1965self} are Schrödinger like, single particle differential equations that can be mapped to the original many-electron Schrödinger equation. DFT offers a balance between accuracy and computational complexity making numerical modeling of realistic materials and devices possible.

TDDFT calculations are typically implemented by either using 
linear response methods \cite{marques2012fundamentals,casida1995time} or a time propagation of the KS orbitals (an approach referred to as real-time TDDFT) \cite{chelikowsky1994higher,marques2012fundamentals,castro2004excited}.

The calculation of large nano-structures, even for the ground state DFT, can be computationally expensive. Moreover, since practical calculations are usually applicable only to small systems,  electromagnetic retardation effects are traditionally neglected in TDDFT calculations.
Neglecting the retardation effects is justified when the system of interest is significantly smaller than the wavelength of the electromagnetic field, but it may lead to incorrect system response behavior for large systems whose size is comparable with the wavelength. 

An approach to compute the electromagnetic fields in large systems is to partition a system into sub-systems \cite{wesolowski1993frozen, krishtal2015subsystem,krishtal2015subsystem2} or to split the computational domain for cases of separable electron densities \cite{shapira2021efficient}. However, existing formulations implementing these ideas for TDDFT do not include time-retarded potentials. 

The inclusion of retardation effects, using a multi-scale scheme, was suggested by A. Yamada and K. Yabana \cite{yamada2019multiscale} who proposed solving the Maxwell's differential equations on a macro scale together with solving the TDDFT equations on a micro scale. Another approach for the inclusion of retardation effects is to use  the equivalent integral expressions for the electromagnetic retarded potentials with the Lorenz gauge \cite{gabay2020lorenz}. The use of this integral approach has several advantages, such as numerical stability, automatic outgoing wave conditions, computational efficiency, and flexibility. 

In this work, we extend the integral approach~\cite{gabay2020lorenz} for systems with separable electron densities, which allow partitioning of the computational domain. An example of such systems can comprise a set of well separated molecules or nano-devices. Since the molecules wave functions do not overlap, their only interaction is electromagnetic. The partitioning of the domain into sub-domains allows much better scaling with the total system size. In systems with large distances between molecules, the partitioning scheme eliminates the empty spaces and leads to a significant speedup.
 We present the benefits of using the Lorenz gauge and  demonstrate the domain partitioning scheme by calculating the retardation effects on pairs of molecules, %which are placed at various%
spaced by various distances comparable to the electromagnetic wavelength. 

\section{Formulation}

We consider a system which can be divided into a set of sub-domains. The density in each sub-domain is spatially separated from the other sub-domains, such that the inter domain interaction is only via electromagnetic potentials. Because of the large spatial separation, the electromagnetic interactions can have large time delays, i.e., electromagnetic retardation effects can become important. Such separable systems represent a number of important cases, for example an array of spatially separated molecules or nano-devices. 

The conventional formulation of TDDFT uses the Coulomb gauge and neglects the magnetic fields. Therefore, the Hartree term is included without retardation. The electric field is typically added as a scalar potential in what is called the length gauge \cite{ullrich2014brief,pemmaraju2018velocity}. However, in electrically large systems it becomes important to include retardation effects, which can be achieved by using the full Coulomb or Lorenz gauge. A formally correct description, which is gauge invariant, is found in time-dependent current density functional theory (TDCDFT) \cite{vignale2004mapping}. In this formalism, the dynamics of a system can be described by a set of single-particle Schr\"odinger like equations for the KS orbitals ${\psi}_{n}$ (atomic units):
\begin{align}\label{TDCDFT KS}
&{i}\frac{\partial}{\partial{t}}{\psi}_{n}\left({\bf{r}},t\right)={\hat{H}}{\psi}_{n}\left({\bf{r}},{t}\right)=
\notag\\&\left[\frac{1}{2}\left(\frac{\nabla}{i}-\frac{q_e}{c}{\textbf{A}}_{s}\left[{\textbf{j}}\right]\left({\bf{r}},{t}\right)\right)^{2}+{v}_{s}\left[{\rho}\right]\left({\bf{r}},{t}\right)\right]\times \notag\\&{\psi}_{n}\left({\bf{r}},{t}\right).
\end{align}
Here, $\hat{H}$ is the dynamic Hamiltonian, $q_e=-1$ since we use atomic units, $c$ is the speed of light in free space, $\rho$ and $\textbf{j}$ are the charge density and current density, respectively, while $\textbf{A}_s$ and $v_s$ are the corresponding effective vector and scalar potentials, respectively. The potentials contain classical phenomena as well as the many-body quantum effects described by exchange-correlation potentials (see appendix \ref{appendix:A} for details). 
Neglecting the exchange-correlation vector part in TDCDFT, it is possible to define the full Coulomb and Lorenz gauges within TDDFT as an approximation~\cite{gabay2020lorenz}.

The Lorenz gauge has an advantage of avoiding the need for a projection scheme of the current density \cite{vignale2004mapping} which is required when using the Coulomb gauge. The formulation for the induced scalar and vector potentials in the Lorenz gauge is described in appendix \ref{appendix:B}.

\begin{figure}
\centering
\includegraphics[width=0.5\textwidth]{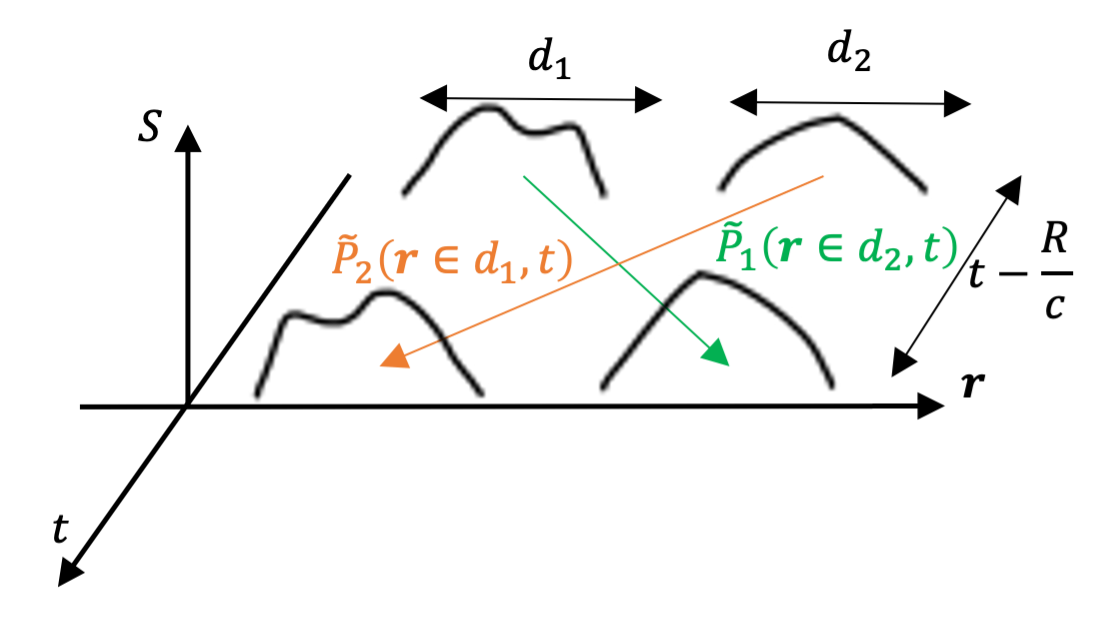}
\caption{A graphic representation of a system with two sub-domains. Each sub-domain induces a delayed scalar and vector potential on the other sub-domain.}
\label{fig:sub domain}
\end{figure}

Let $S\in  \{{q_e}\rho,{q_e}\textbf{j}/c\}$ denote the four-vector source density and $P\in  \{v_{\mathrm{ind}},\textbf{A}_{\mathrm{ind}}\}$ denote the four-vector induced potential. The governing integral expressions for the induced scalar and vector potentials via the Lorenz gauge, which naturally contains retardation, are:
\begin{equation} \label{4v potential}
    {{P}}(\textbf{r},t)=\int{\frac{S(\textbf{r}',t-\frac{R}{c})}{R}}d\textbf{r}',
\end{equation}
where $R=|\textbf{r}-\textbf{r}'|$ is the distance between the  source and observation points. In a case where the domain consists of a number of spatially separated densities, we can apply a multi-domain scheme by splitting the charge and current  densities into separate sub-domains. The resulting induced four-vector potential is obtained as a sum over the sub-domain contributions:
\begin{equation}\label{multi domain TDCDFT lorenz induced 4 vector convulotion}
  P(\textbf{r},t)=\sum_{n=1}^{N_{\mathrm{dom}}}\tilde P_{n}(\textbf{r},t),
\end{equation}
where $N_{\mathrm{dom}}$ is the number of distinct sub-domains, and $\tilde P_{n}$ represents the scalar and vector potentials induced by the sources of the $n$th sub-domain $d_n$. A demonstration of this procedure is depicted in Fig. \ref{fig:sub domain}. 
The potentials induced by each sub-domain can be evaluated similarly to Eq. (\ref{4v potential}):
\begin{equation} \label{4v potential multi}
    {\tilde{P}}_{n}(\textbf{r},t)=\int_{{d_n}}{\frac{S(\textbf{r}',t-\frac{R}{c})}{R}}d\textbf{r}'.
\end{equation}

The induced potentials due to each sub-domain are calculated by the process handling this sub-domain for all the other sub-domains. Hence, in each time step of the TDDFT propagation, each sub-domain calculates its potential contribution at all other sub-domains as well as collects the calculated potential contributions from all other sub-domains (see Fig. \ref{fig:multi domain time domain scheme} for an illustration). This method has a quadratic computational scaling with respect to the number of sub-domains and its computational complexity is independent of the distance between sub-domains.

Since the induced potentials are evaluated at each time step of the propagation of the TDKS orbitals, and the retardation requires information from the past times, this process consumes a large amount of memory, in addition to being computationally intensive. In the next section we present a discrete formulation for evaluating the potential of Eqs. (\ref{multi domain TDCDFT lorenz induced 4 vector convulotion}) and (\ref{4v potential multi}) and its efficient implementation. To minimize the memory consumption we define the minimal number of necessary time steps, which are needed to store the past calculations. To reduce the computational complexity of  Eq. (\ref{4v potential multi}), we use an efficient FFT-based method.

\section{Implementation and Discrete Space-Time Approach}
Equation \eqref{4v potential multi} can be reformulated as the temporal convolution:
\begin{equation}\label{multi domain TDCDFT lorenz induced 4 vector convulotion per domain}
  {\tilde{P}}_{n}(\textbf{r},t)=\int_{d_n} G(R,t)\ast {\tilde{S}}(\textbf{r}',t)d\textbf{r}'.
\end{equation}
Here, the asterisk denotes a time domain convolution and $G$ is the time domain free-space Green's function:
\begin{equation}\label{multi domain time domain Greens' function}
  G(R,t)= \frac{\delta(t-\frac{R}{c})}{R}
\end{equation}
Since the domains are assumed to be non-overlapping, no specific treatment for the singular point of the Green's function is required, when the potential is computed in a sub-domain different from the source one. A smoothed Green's function for the case of potential calculations within a domain is described in \cite{gabay2017optimizing}.

To calculate the four-vector potential in the discrete space-time domain, the sources (charge and current densities) are sampled uniformly in space using $N_{{\mathrm{g}},n}=N_{x,n}\times N_{y,n}\times N_{z,n}$ grid points with a grid step of $h$, and in time with a time step of $\Delta t$. The total four-vector source density is approximated using the interpolation in time
\begin{equation}\label{space-time sampled four vector sources}
  S(\textbf{r}_{k},t)\cong S(\textbf{r}_{k},0)+\sum_{l=1}^{N_t}\Delta S(\textbf{r}_{k},l\Delta t)T(t-l\Delta t),
\end{equation}
where $\textbf{r}_{k}=(x_k,y_k,z_k)$ for $1\leq{k}\leq{N_{\mathrm{g,n}}}$, $\Delta S(\textbf{r}_{k},l\Delta t)$ is the difference between the sampled four-vector source density at time $t=l\Delta t$ and the initial density (ground state), i.e., $\Delta S(\textbf{r}_{k},l\Delta t)= S(\textbf{r}_{k},l\Delta t)-S(\textbf{r}_{k},0)$. Here $N_t$ is the number of time samples up to time $t$, i.e, $N_t=\lceil {t}/{\Delta t}\rceil$, where $\lceil\cdot\rceil$ denoted a ceiling function and $T(\cdot)$ is a temporal triangle basis function used for interpolation in the predictor-corrector propagation scheme \cite{mundt2007photoelectron,mundt2009real}. 
Substitution of Eq. (\ref{space-time sampled four vector sources}) into (\ref{multi domain TDCDFT lorenz induced 4 vector convulotion per domain}) results in a procedure to compute the four-vector potential generated by domain $n$ at domain $s$. The four-vector potential is obtained as:
\begin{align}\label{multi domain TDCDFT lorenz space-time sampled induced 4 vector convulotion}
  {\tilde{P}}_{n}(\textbf{r}_{k},t)&=\sum_{\textbf{r}_{k'}\in{d_n}} h^3G(R_{k,k'},t)\ast\bigg[{S}(\textbf{r}_{k'},0)+\nonumber\\  &\sum_{l=1}^{N_t}\Delta {S}(\textbf{r}_{k'},l\Delta t)T(t-l\Delta t)\bigg],
\end{align}
where $R_{k,k'}=|{\bf{r}}_{k}-{\bf{r}}_{k'}|.$% and ${S}^{(n)$ is the four-vector source density of the $n_{th}$ domain.

Unfortunately, the direct calculation via eq. (\ref{multi domain TDCDFT lorenz space-time sampled induced 4 vector convulotion}) is computationally  expensive with the complexity scaling as $\mathcal{O}(N_tN_{\mathrm{g},n}^2)$. These calculations can be accelerated by FFT technique over the spatial variable. We use the distributive property of the convolution to write:
\begin{flalign}\label{multi domain TDCDFT lorenz space-time sampled simplify 1}
{\tilde{P}}_{n}(\textbf{r}_{k},t)&=\tilde{P}_{n}(\textbf{r}_{k},0)+&&\\\nonumber
            & \sum_{r_{k'}\in{d_n}}\sum_{l=1}^{N_t} h^3G(R_{k,k'},t)\ast&&\\\nonumber
            & \ \ \ \ \ \ \ \ \ \ \ \space \left[\Delta {S}(\textbf{r}_{k'},l\Delta t)T(t-l\Delta t)\right],
\end{flalign}
where $P(\textbf{r}_{k},0)$ is the ground-state four-vector potential that can be calculated using the static Poisson-equation solver or FFT based superposition \cite{gabay2017optimizing}. The second term of Eq. (\ref{multi domain TDCDFT lorenz space-time sampled simplify 1}) can be rewritten using the associativity property of the convolution as:
\begin{align}\label{retardation space-time sampled}
  &G(R_{k,k'},t)\ast\left[\Delta {S}(\textbf{r}_{k'},l\Delta t)T(t-l\Delta t)\right]=\nonumber \\&\Big[G(R_{k,k'},t)\ast T(t-l\Delta t)\Big]\Delta{S}(\textbf{r}_{k'},l\Delta t).
\end{align}

We now define the space-time sampled Green's function adequate for the problem as:
\begin{align}\label{retardation space-time sampled green interpolation}
 G_{l-l'}[k,k']&=h^3G(R_{k,k'},t)\ast T(t-l'\Delta t)|_{t=l\Delta t}\nonumber \\&=\frac{T\left((l-l')\Delta t-\frac{|\textbf{r}_k-\textbf{r}_{k'}|}{c}\right)}{|\textbf{r}_k-\textbf{r}_{k'}|}.
\end{align}
As a result, the total four-vector potential induced by multi-domains is obtained via
\begin{align}\label{retardation space-time sampled lorentz four vector potential}
  {{P}}(\textbf{r}_{k},&l\Delta t)=\sum_{n=1}^{N_{\mathrm{dom}}}\bigg[ \bigg. \tilde{P}_{n}(\textbf{r}_{k},0)+\nonumber
  \\&\sum_{\textbf{r}_{k'}\in{d_n}}\sum_{l'=1}^{l}G_{l-l'}[k,k']\Delta{S}(\textbf{r}_{k'},l'\Delta t) \bigg. \bigg].
\end{align}
It should be noted that the sum over $l^\prime$, although indicated to be over all past times, actually is only over times that are determined by the retardation relevant to the sub-domain sizes and distances. This point is explained in detail in the following section.

\section{Computational Complexity Analysis}
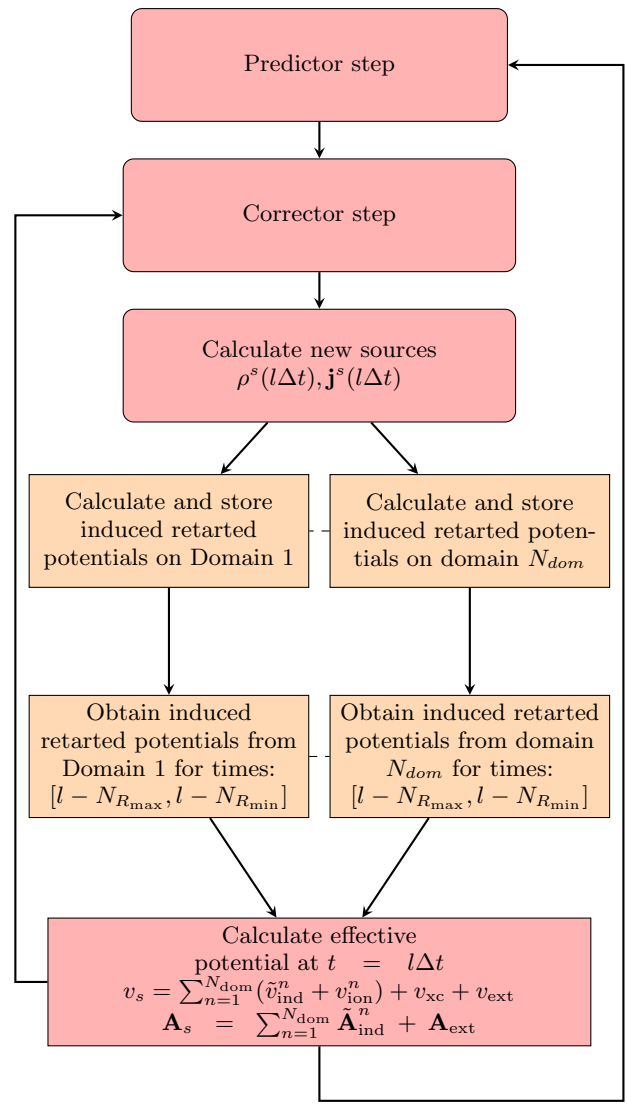
\begin{figure}
\tikzstyle{startstop} = [rectangle, rounded corners, minimum width=5cm, minimum height=1.5cm,text centered, draw=black, fill=red!30]
\tikzstyle{io} = [trapezium, trapezium left angle=70, trapezium right angle=110, minimum width=5cm, minimum height=2cm,text width=5cm, text centered, draw=black, fill=blue!30]
\tikzstyle{process} = [rectangle, minimum width=3cm, minimum height=1.5cm,text width=5cm, text centered, draw=black, fill=orange!30]
\tikzstyle{decision} = [diamond, minimum width=5cm, minimum height=1cm, text centered, draw=black, fill=green!30]
\tikzstyle{arrow} = [thick,->,>=stealth]
{\centering\begin{tikzpicture}[node distance=2cm,auto]
\node (start) [startstop] {Predictor step};
\node (in1) [process, below of=start, fill=red!30, rounded corners] {Corrector step};
\node (pro1) [process, below of=in1,fill=red!30, rounded corners] {Calculate new sources $\rho^s(l\Delta t),{\bf{j}}^{s}(l\Delta t)$};
\node (calc ind pot 1) [process, below of=pro1, xshift=-2cm,node distance=2.2cm,text width=3.5cm] {Calculate and store induced retarted potentials on Domain 1 % \\ $\tilde{P}^s(\textbf{r}_{k}\in{d^1},l\Delta t)=\sum_{\textbf{r}_{k}'\in d^s}G_{l-l'}[k,k']\tilde{S}^s(\textbf{r}_{k'},l'\Delta t)$%
};
\node (calc ind pot N) [process, below of=pro1, xshift=2cm,node distance=2.2cm,text width=3.5cm] {Calculate and store induced retarted potentials on domain $N_{dom}$ \\ %$\tilde{P}^s(\textbf{r}_{k}\in{d_n},l\Delta t)=\sum_{\textbf{r}_{k}'\in d^s}G_{l-l'}[k,k']\tilde{S}^s(\textbf{r}_{k'},l'\Delta t)$%
};
\node (get ind pot 1) [process, below of=calc ind pot 1,node distance=3cm,text width=3.5cm] {Obtain induced retarted potentials from Domain 1 for times:\\$[l- N_{R_\mathrm{max}},l- N_{R_\mathrm{min}}]$ %$l-\tilde{l}, \quad \tilde{l} \in [Ng_{\mathrm{max}}^{(S,1)},Ng_{\mathrm{min}}^{(S,1)}]$\\$\tilde{P}^s(\textbf{r}_{k}\in{d^s},l\Delta t)=\sum_{l-\tilde{l}}\tilde{P}^N(\textbf{r}_{k}\in{d^s},l-\tilde{l})$%
};
\node (get ind pot N) [process, below of=calc ind pot N,node distance=3cm,text width=3.5cm] {Obtain induced retarted potentials from domain $N_{dom}$ for times:\\$[l- N_{R_\mathrm{max}},l- N_{R_\mathrm{min}}]$%\\ $l-\tilde{l}, \quad \tilde{l} \in [Ng_{\mathrm{max}}^{(S,N)},Ng_{\mathrm{min}}^{(S,N)}]$\\$\tilde{P}^s(\textbf{r}_{k}\in{d^s},l\Delta t)=\sum_{l-\tilde{l}}\tilde{P}^N(\textbf{r}_{k}\in{d^s},l-\tilde{l})$%
};
\node (out1) [process, below of=get ind pot N, xshift=-2cm, fill=red!30,node distance=3cm,text width=7cm] {Calculate effective\\ potential at $t=l\Delta t$ \\${ v_s=\sum_{n=1}^{N_{\mathrm{dom}}}(\tilde{v}_{\mathrm{ind}}^n+v_{\mathrm{ion}}^n)+v_{\mathrm{xc}}+v_{\mathrm{ext}} }$\\ $\textbf{A}_{s}=\sum_{n=1}^{N_{\mathrm{dom}}}{\tilde{\textbf{A}}^n_{\mathrm{ind}}}+\textbf{A}_{\mathrm{ext}}$};
%\node (stop) [startstop, below of=out1] {Stop};
\draw [arrow] (start) -- (in1);
\draw [arrow] (in1) -- (pro1);
\draw [arrow] (pro1) -- (calc ind pot 1);
\draw [arrow] (pro1) -- (calc ind pot N);
\draw [arrow] (calc ind pot 1) -- (get ind pot 1);
\draw [arrow] (calc ind pot N) -- (get ind pot N);
\draw [arrow] (get ind pot 1) -- (out1);
\draw [arrow] (get ind pot N) -- (out1);
\draw [arrow] (out1)  -- ++(-115pt,0pt) |- (in1) ;
\draw [arrow] (out1)  -- ++(0pt,-45pt)-- ++(+115pt,0pt) |- (start) ;
\draw[dashed] (calc ind pot N) -- (calc ind pot 1);
\draw[dashed] (get ind pot 1) -- (get ind pot N);
\end{tikzpicture}}
\caption{Propagation scheme modified to integrate multi-domain calculations. Each domain produces the current time induced potentials on all the domains and obtains the retarded potentials from other domains (orange rectangle processes). Then these potentials are accumulated for a total effective potential, and propagated forward in time (red rectangle processes). The predictor-corrector scheme is used as in \cite{mundt2009real,mundt2007photoelectron}, where the induced retarded potentials are injected into the total effective potential in each step.}
\label{fig:multi domain time domain scheme}
\end{figure}

This section describes the efficiency and limitations of the multi-domain scheme. We start with a detailed analysis of the induced retarded vector potential calculations. Then, we describe the full TDDFT calculations (see Fig. \ref{fig:multi domain time domain scheme}) and the integration of the multi-domain scheme with the time propagation algorithm. We also analyze the computational complexity of the multi-domain scheme for several scenarios.

Equation (\ref{retardation space-time sampled lorentz four vector potential}) is a discrete convolution procedure with respect to the temporal index $l'$. Its direct evaluation is expensive.
By utilizing FFT-based techniques  \cite{eastwood1979remarks,cerioni2012efficient}, which use a 3D FFT scheme along each slice of time, it is possible
to reach a computational complexity of $O(N_tN_{\mathrm{g}}\mathrm{log}N_{\mathrm{g}})$ operations. This computational complexity is in fact lower even for a single domain since one needs to take into account only the maximal retardation time backward as the interpolation function is zero outside of the domain. Hence, the actual complexity is $O(N_RN_{\mathrm{g}}\mathrm{log}N_{\mathrm{g}})$, where $N_R={D}/{c\Delta t}$
is the maximal number of time steps required to propagate over the domain and $D$ is the maximal distance between any two points inside the domain.

\begin{figure}
\centering
\includegraphics[width=0.5\textwidth]{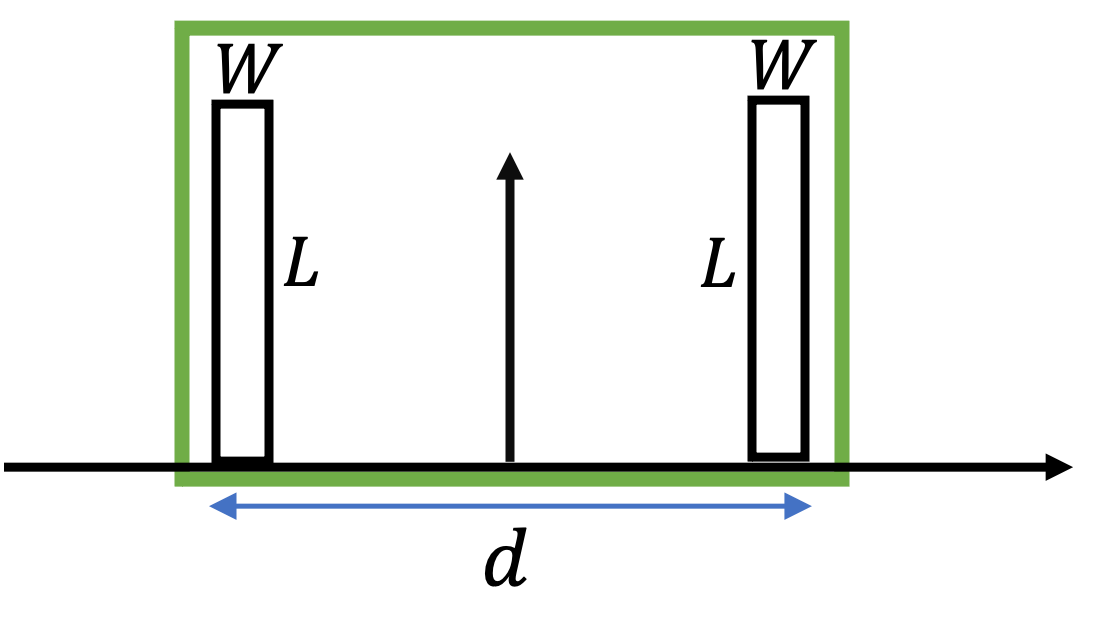}
\caption{A setup of two elongated sub-domains of a size ${W}\times{W}\times{L}$, with a dominant axis ($L>>W$). The sub-domains are set in parallel to each other with a distance of $d$ with respect to their horizontal axis. The green shaped box represents the size of a single large domain that contains both sub-domains (it is slightly larger only for illustration purposes).}
\label{fig:domains complexity}
\end{figure}

\begin{figure}
\centering
\includegraphics[width=0.5\textwidth]
{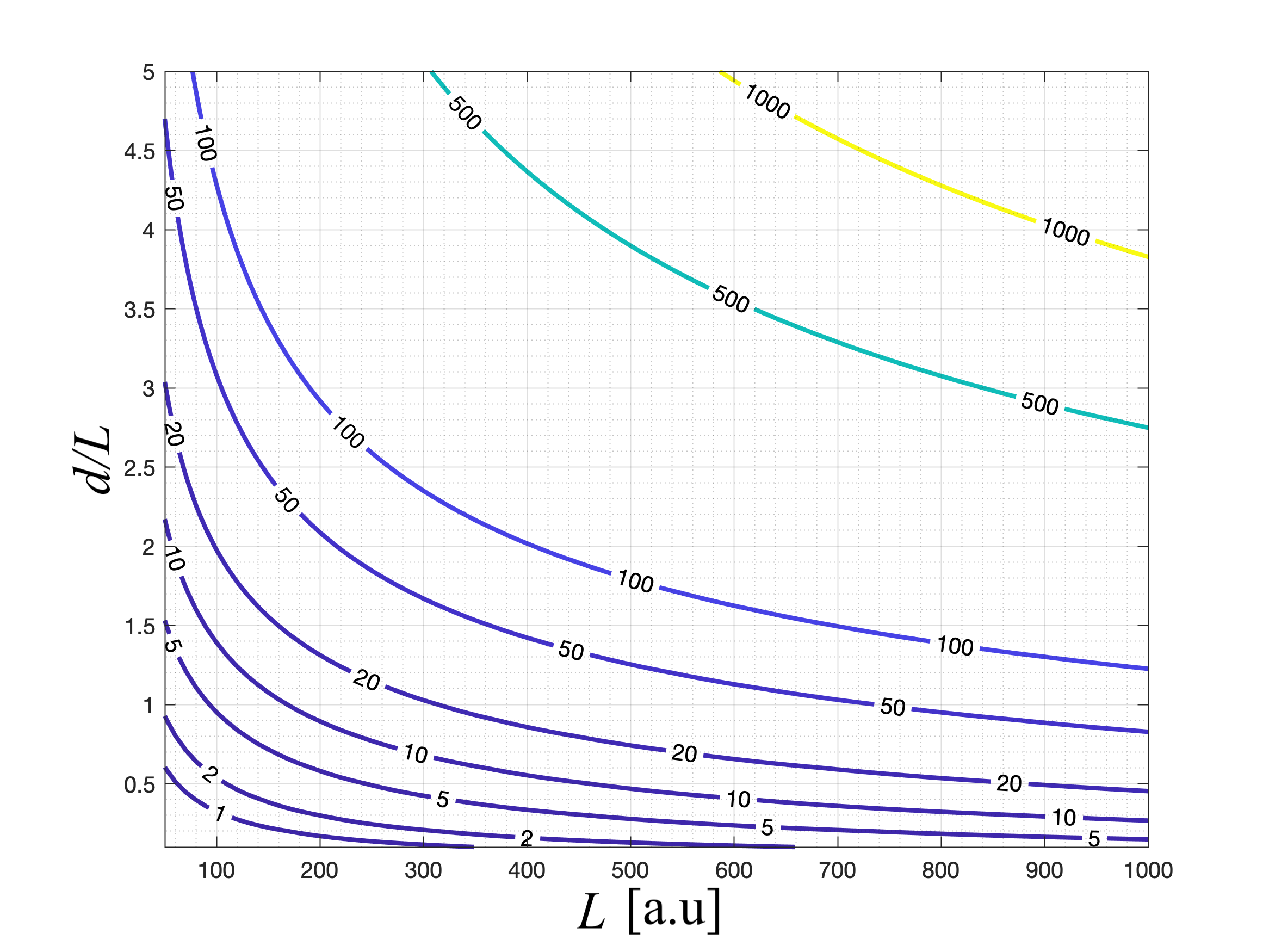}
\caption{Computational complexities ratio between the single domain and the multi-domain i.e $speedup$. The domains' parameters are $W=10$ a.u., $L$ varies from $50$ a.u. to $1000$ a.u. and $d$ according to the y axis.}
\label{fig:complexity ratio general}
\end{figure}

By using the multi-domain method, the actual size of time slices needed for the estimation of the radiation from the $n$th sub-domain to the $s$th sub-domain is reduced and the related computational complexity is $O(\Delta N_{{R,ns}}{N}_{{\mathrm{g},n}}\mathrm{log}{N}_{{\mathrm{g},n}})$, where ${N}_{\mathrm{g},n}$ is the number of grid points in the $n$th sub-domain (assuming ${N}_{\mathrm{g},n}\simeq{N}_{\mathrm{g},s}$), $\Delta N_{R,ns}={\Delta D_{ns}}/{c\Delta t}$ is the number of past steps needed for the current potential evaluation, and $\Delta D_{ns}=[\mathrm{max}(|\textbf{r}_i-\textbf{r}_j
|)-\mathrm{min}(|\textbf{r}_i-\textbf{r}_j
|) | \textbf{r}_i\in d_s,\textbf{r}_j\in d_n]$ is the difference between the maximal and minimal distance between the domains. In other words, $\Delta N_R$ is the difference between the maximal and minimal time steps needed to propagate between any two points on each domain. Hence, the partitioning of the large domain into sub-domains reduces the numbers of both the grid points and retardation time steps needed for each sub-domain. 
To compare between the complexities of multi-domain and single large domain setups, we first examine a configuration of two identical elongated sub-domains (see Fig. \ref{fig:domains complexity}). For simplicity, we assume that each sub-domain has a long axis of size $L$, and transversal cross section of size ${W}\times{W}$  ($W\ll L$). The sub-domains are located at distance $d$ from each other along the transversal axis with ${W}\ll{d}$. The number of grid points in each sub-domain is $\tilde{N}_\textnormal{g}={W^2}{L}/h^3={{W'}^2}{{L'}}$, with $W'=W/h$ , $L'=L/h$, whereas the number of grid points of the whole large domain containing both sub-domains is $N_\textnormal{g}={W'}{d'}{L'}$, with $d'=d/h$. 
The number of retardation time steps ($\Delta N_R$) that is needed for the multi-domain scheme is a function of $\Delta D$. For the suggested geometry, it can be calculated as the difference between the diagonal of the full domain ($\sqrt{{d'^2}+{L'^2}+{W'^2}}$) and the horizontal distance between the sub-domains ($d'-2W'$).
Finally, the total computational complexity of the multi-domain setup  is composed of the computation of the retarded potential between the distinct domains and the computation of the retardation inside each sub-domain. The total computational complexity ratio of the four-vector potential between the multi-domain and a full single domain for the suggested configuration is given by:
\begin{align}
\label{computetional complexity ratio}
&{\textnormal{Speedup}}^{-1} = \nonumber\\ &\frac{2\big(\sqrt{d'^2+L'^2+W'^2}\big)W'^2L'\textnormal{log}(W'^2L')}{\big(\sqrt{d'^2+L'^2+W'^2}\big)W'd'L'\textnormal{log}(W'd'L')} 
\nonumber\\
&
-\frac{2\big((d'-2W')\big)W'^2L'\textnormal{log}(W'^2L')}{\big(\sqrt{d'^2+L'^2+W'^2}\big)W'd'L'\textnormal{log}(W'd'L')} 
\nonumber\\
&+\frac{2\big(\sqrt{2W'^2+L'^2})\big)W'^2L'\textnormal{log}(W'^2L')}{\big(\sqrt{d'^2+L'^2+W'^2}\big)W'd'L'\textnormal{log}(W'd'L')}
\end{align}
The Speedup depends on the ratio between $L$ and $d$ (assuming $W\ll L$, $W\ll d$). In the regime of $d \gg L$, the computational complexity of the multi-domain setup is independent of the value of $d$ and the efficiency of the multi-domain method is dominant.
Figure \ref{fig:complexity ratio general} shows the speedup of the multi-domain method as a function of $d/L$ for different $L$. The domain width was chosen to be $W=20$ a.u. and the domain length varied between $L=10$ a.u. to $L=1000$ a.u.

It is important to emphasize that although we only need $\Delta N_R$ past densities to calculate the current time potentials, it is necessary to store the densities at all times in the range between the current time and the maximal number of time steps needed to propagate between domains ($[N_t- N_{R_\mathrm{max}},N_t]$) for later use to avoid their re-computation (see Fig. \ref{fig:multi domain time domain scheme} for the full scheme). Although this may consume a large amount of memory, it is not necessary to store them in random access memory (RAM), but rather-on the storage drive, which is a significant advantage.   
The main bottleneck of the TDDFT calculations is the computation of the retarded four-vector potential and the memory requirements at each time step. Another major computational bottleneck is the the application of the Hamiltonian ($\hat{H}\psi_n$) in the time propagation and the evaluation of the current density and current.
A predictor-corrector \cite{mundt2009real,mundt2007photoelectron} %[\ref{propagator taylor expansion}]%
scheme is adopted for the time propagation, and to ensure properly converged dynamics. The application of the Hamiltonian at each time step,  % Eq. (\ref{predictor step PARSEC}-\ref{corrector step PARSEC})
typically requires $O(N_{\mathrm{g}}N_o)$ operations, where $N_o$ is the number of orbitals. Within the suggested method, the computational complexity of this step is reduced by a factor of at least the number of sub-domains to $O(N_{\mathrm{g}}N_o/N_{\mathrm{dom}})$.
The evaluation of the density and current also requires $O(N_{\mathrm{g}}N_o)$ operation (see \ref{appendix:A}), but is usually much faster. This computation complexity is also reduced to $O(N_{\mathrm{g}}N_o/N_{\mathrm{dom}})$ in the multi-domain scheme, this is because the the current contribution of each orbital's is calculated in the sub-domain instead of the full domain.
Hence, the multi-domain scheme provides an effective procedure to reduce the computational burden for spatially separable setups. 
Moreover, this method is suitable for parallelization, which is efficient because every domain only needs to retrieve information from the previous calculations of other domains and prepare information for the future use by other domains. The information transfer can be done efficiently using file system methods assuming that all the processes have access to a shared disk or using a fast communication with a file transfer protocol such as the MQTT \cite{hunkeler2008mqtt}. The whole TDDFT sequence modified with the multi-domain scheme is illustrated in Fig. \ref{fig:multi domain time domain scheme}.
%\subsection{Multiline equations}
\section{Results}
We now demonstrate the use of our scheme to calculate excitation for a pair of C$_{12}$H$_4$ molecules (see Figs. \ref{fig:dimers} and \ref{fig:c12h4}). We also present additional results for N$_2$ in appendix \ref{appendix:c}. A few more examples for the multi-domain scheme for the calculation of the ground state properties were presented in our previous work \cite{shapira2021efficient}.
We implemented the method within the Bayreuth version~\cite{mundt2009real,mundt2007photoelectron} of the PARSEC real-space software package \cite{chelikowsky1994higher,kronik2006parsec}. We used norm conserving pseudopotentials \cite{troullier1991efficient} of s/p/d cutoffs: C: 1.6/1.6, H: 1.39, Cl: 1.74/1.74/1.74, N: 1.5/1.5, O: 1.29/1.29. Furthermore, we used the adiabatic local density approximation (ALDA) functional for the exchange-correlation scalar potential. The domain size of each molecule was a box of 20$\times$20$\times$50 a.u. with a grid spacing of $0.4$ a.u.
We used absorbing boundary conditions (ABC) with a boundary layer of $5$ a.u. on each spatial dimension and exponential damping coefficient of $0.025$ a.u. The electronic structures were evolved in time for $30$ femtoseconds with a time step of $1$ attosecond. These parameters were all specified to ensure numerically correctness of the model.  Fourth-order Taylor expansion and a predictor-corrector scheme are used to propagate explicitly the KS orbitals \cite{mundt2009real,mundt2007photoelectron}.\\
The applied external electric field used to perturb the electronic structure is linearly polarized along the $\hat{z}$ direction. It takes the form of a sine-squared pulse:
\begin{equation}\label{laser energy}
{\bf{E}}_{\text{ext}}\left({t}\right) = {A}_{s}{\sin}^{2}\left({\omega}_{\text{env}}{t}\right){\sin}\left({\omega}{t}\right){\bf{\hat{z}}}.
\end{equation}
Here, ${A}_{s}$, ${\omega}$, ${\omega}_{\text{env}}$ are the intensity, radial frequency, and radial envelope frequency of the pulse, respectively. We assumed that the cross section of each domain is small enough to approximate the field as spatially constant. A laser wavelength of $800$ nm, intensity of $5\times10^{12}$ W/{cm}$^2$ and envelope period of $T_\textnormal{env}=26$ fs was used throughout. 

To demonstrate the multi-domain model and to investigate the coupling and retardation effects we used the following setup. First, two symmetric separate molecules were placed in parallel to the $z$-axis at various distances between them (see Fig. \ref{fig:dimers}).
Then, we applied an external field on one of the domains ("Domain 1") and captured the induced dynamics in the second domain ("Domain 2"). A laser pulse as in Eq. (\ref{laser energy}) was applied as an external field to Domain 1.
\begin{figure}
\centering
\includegraphics[width=0.5\textwidth]{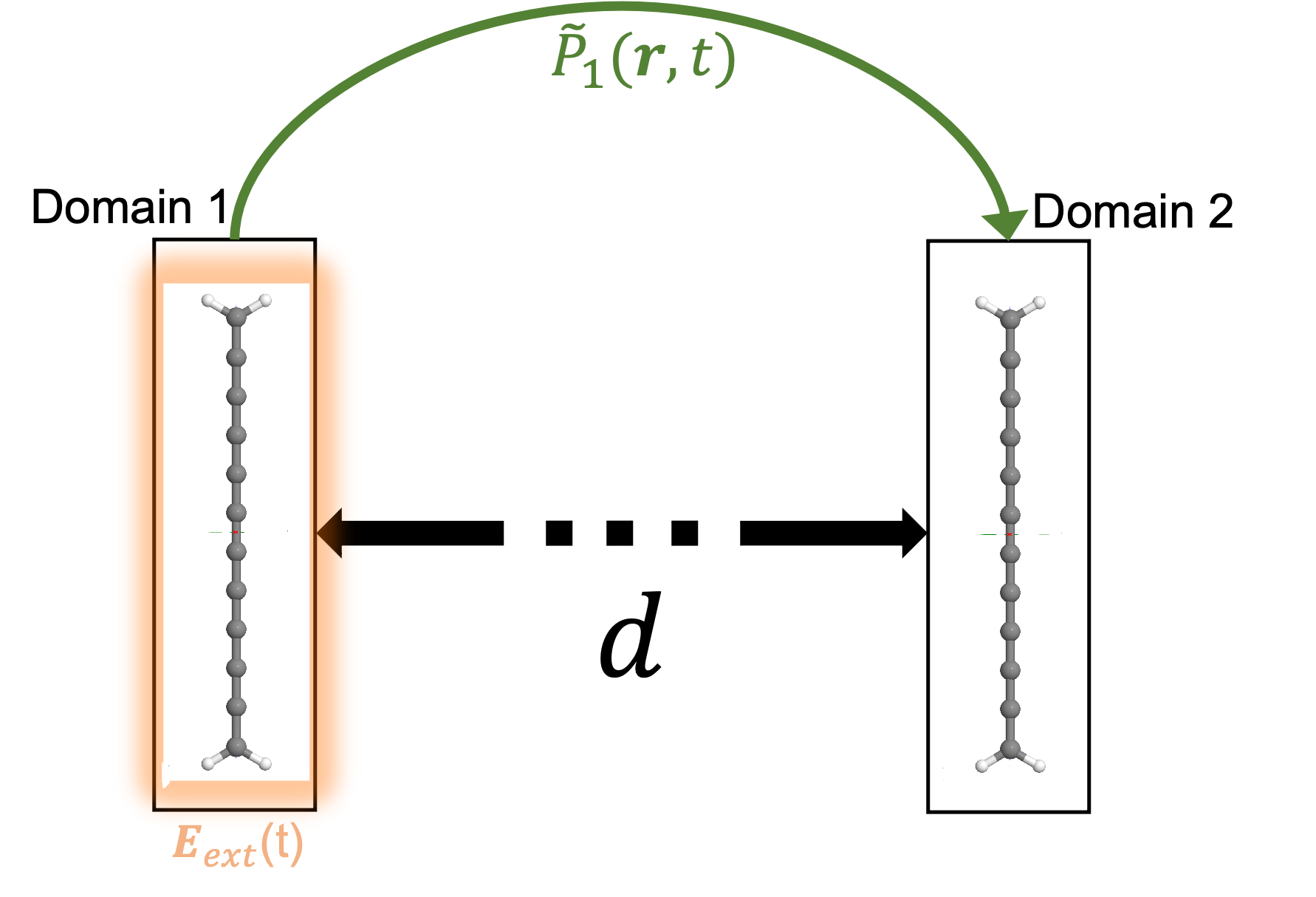}
\caption{The setup used to demonstrate the model. Two sub-domains of C$_{12}$H$_4$ with parallel axes, located at various distances. An external field is applied to Domain 1. Domain 2 is affected by the induced potential of Domain 1.}
\label{fig:dimers}
\end{figure}

When the molecules are far from one another, the behavior of the electromagnetic field in sub-domain 2, originating from sub-domain 1, can be approximated by the field of a %monochromatic oscillating infinitesimal%
Hertzian dipole:
\begin{align}\label{eq: electric field inf dipole}
&{E_{1,z}(t)}=\notag\\&-\frac{{D_{1,z}(t-t_r)}e^{-i\omega(t-t_r)}}{4\pi\epsilon_o}\bigg(\frac{\omega^2}{c^2d}+\frac{i\omega}{cd^2}-\frac{1}{d^3}\bigg)
\end{align}
where $d$ is the distance between the domains, $t_r=d/c$ is the retardation time, 
and $D_{1,z}$ is the amplitude of the time dependent $z$-component dipole moment(assuming the envelope changes slowly with respect to the retardation time). We consider just the $z$ component of the induced field because the molecules we choose are centro-symmetric, hence their polarizability tensor is a diagonal matrix in the chosen geometry. The time dependent dipole moment can be calculated using the density of the single-particle KS orbitals: 
\begin{equation}
\textbf{D}(t)=\int{\textbf{r}\sum_{N_o}|{\psi}_{i}(\textbf{r},t)|^2d\textbf{r}}
\end{equation}
The effect on the dipole moment of Domain 2 can be estimated by the molecule polarizability \cite{griffiths2005introduction}: 
\begin{equation}\label{polarizability as function of electric field}
{D}_{2,z}=\alpha_{z,z}{E}_{1,z}
\end{equation}
where $\alpha_{z,z}$ is the diagonal element of the polarizability tensor of the molecule in the $z$-direction. 
We derived the value of $\alpha_{z,z}$ from a single molecule calculation of the ratio between the induced dipole moment and the laser amplitude in the frequency domain. The approximation in equations \ref{eq: electric field inf dipole}-\ref{polarizability as function of electric field} is valid only when the distance between molecules is large relative to their size, we therefore expect it to agree with the full calculation only at such distances.

We study the response of a 1D carbon carbyne chain in the cumulene form, with the chemical formula of %$\chemfig{H=C=H}$%
C$_{12}$H$_4$ (see Fig. \ref{fig:c12h4}).
\begin{figure}[htb!]
\centering
\includegraphics[width=0.5\textwidth]{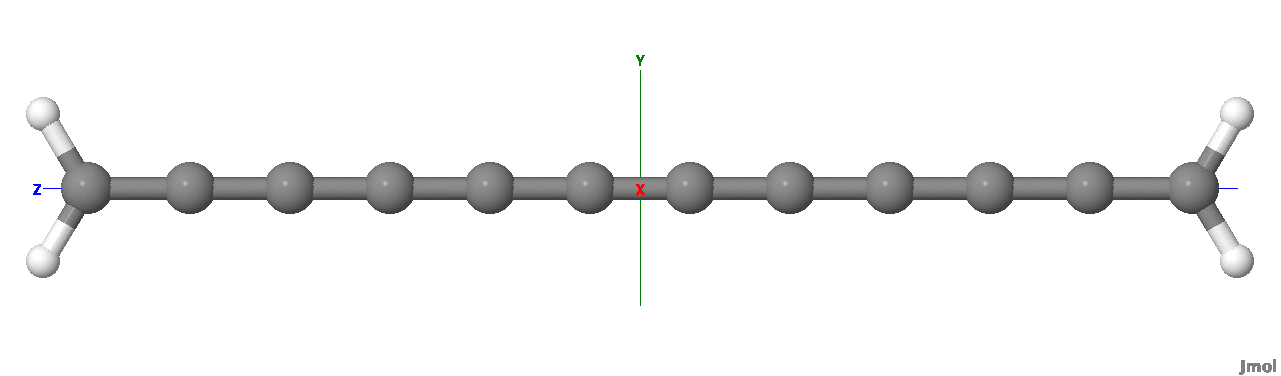}
\caption{single molecule structure of 1D carbon carbyne chain $H_2(=C=)_{12}H_2$, 
the hydrogen atoms (white balls) are placed along the $yz$ plane.
the Carbon atoms (grey balls) are placed along the $z$-axis.}
\label{fig:c12h4}
\end{figure}
The C-C and C-H bond lengths are $2.39$ a.u. and $2.07$ a.u., respectively. This form of carbyne has a zero band gap in the infinite  polymer limit \cite{calzolari2004ab}. This means that for finite molecules we can expect increasingly larger polarizability as they increase in size, hence they can serve as an excellent toy model for nano-antennas. 

The setup included two molecules in two distinct domains that were separated along the $y$-axis. As stated above, we applied a laser as in Eq. (\ref{laser energy}), with linear polarization along the $z$-axis, on Domain 1 and captured the response of Domain 2 (see Fig. \ref{fig:dimers}). The response of Domain 2 is hence purely a result of the molecule-molecule interactions. 

\begin{figure}
\centering
\includegraphics[width=0.5\textwidth]{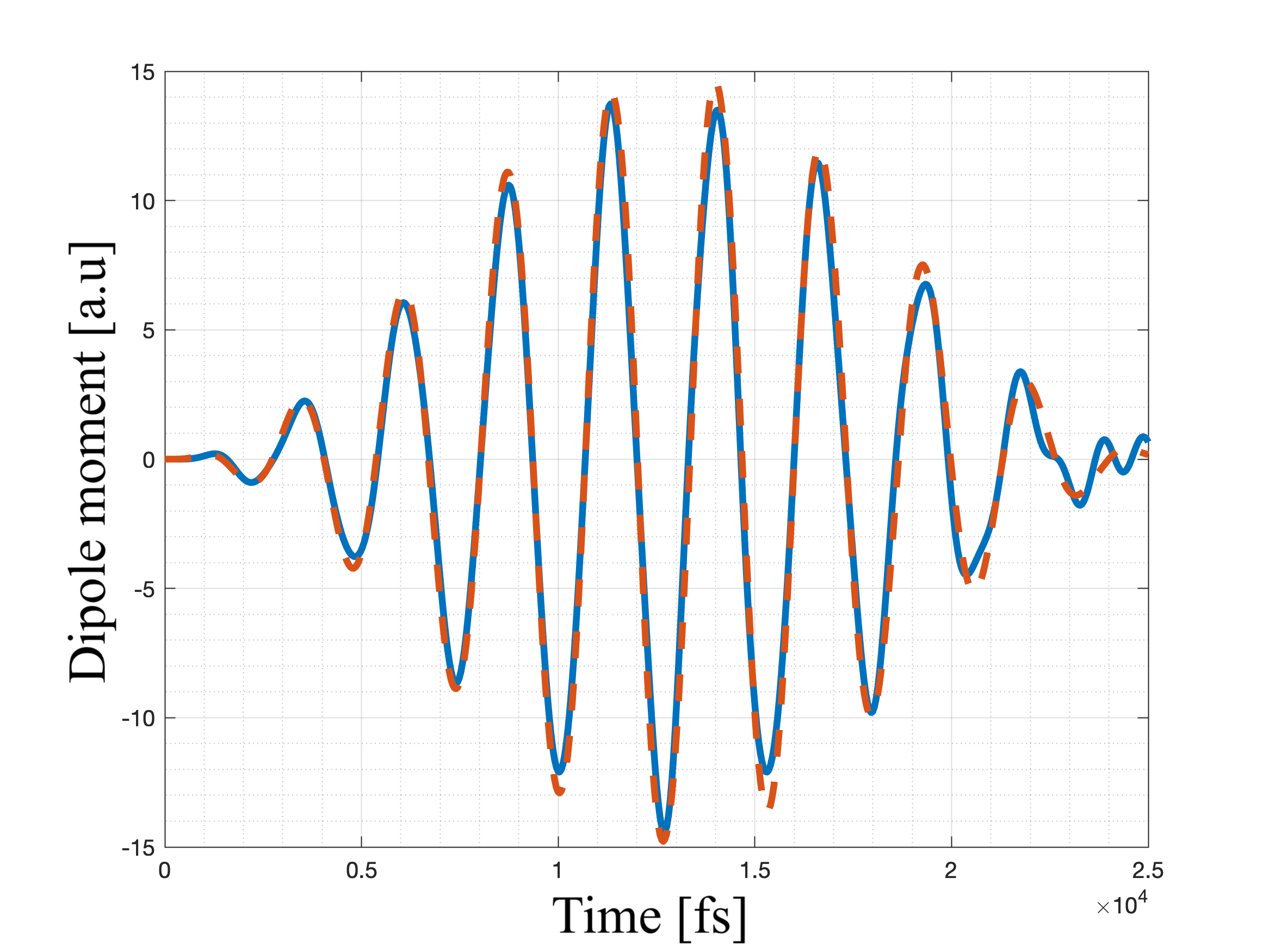}
\caption{Comparison of the dipole moment along the C$_{12}$H$_{4}$ molecules axis ($\bf{\hat{z}}$ direction), with separation of $d$=90 a.u. ($\approx 4.76$ nm) between the domains. in red (dash) the domain with the applied laser (Domain 1). Domain 2 (blue) is normalized with the factor of the dipole response for comparison.}
\label{fig:c12h4_90_td}
\end{figure}

\begin{figure}[h]
\centering
\includegraphics[width=0.5\textwidth]
{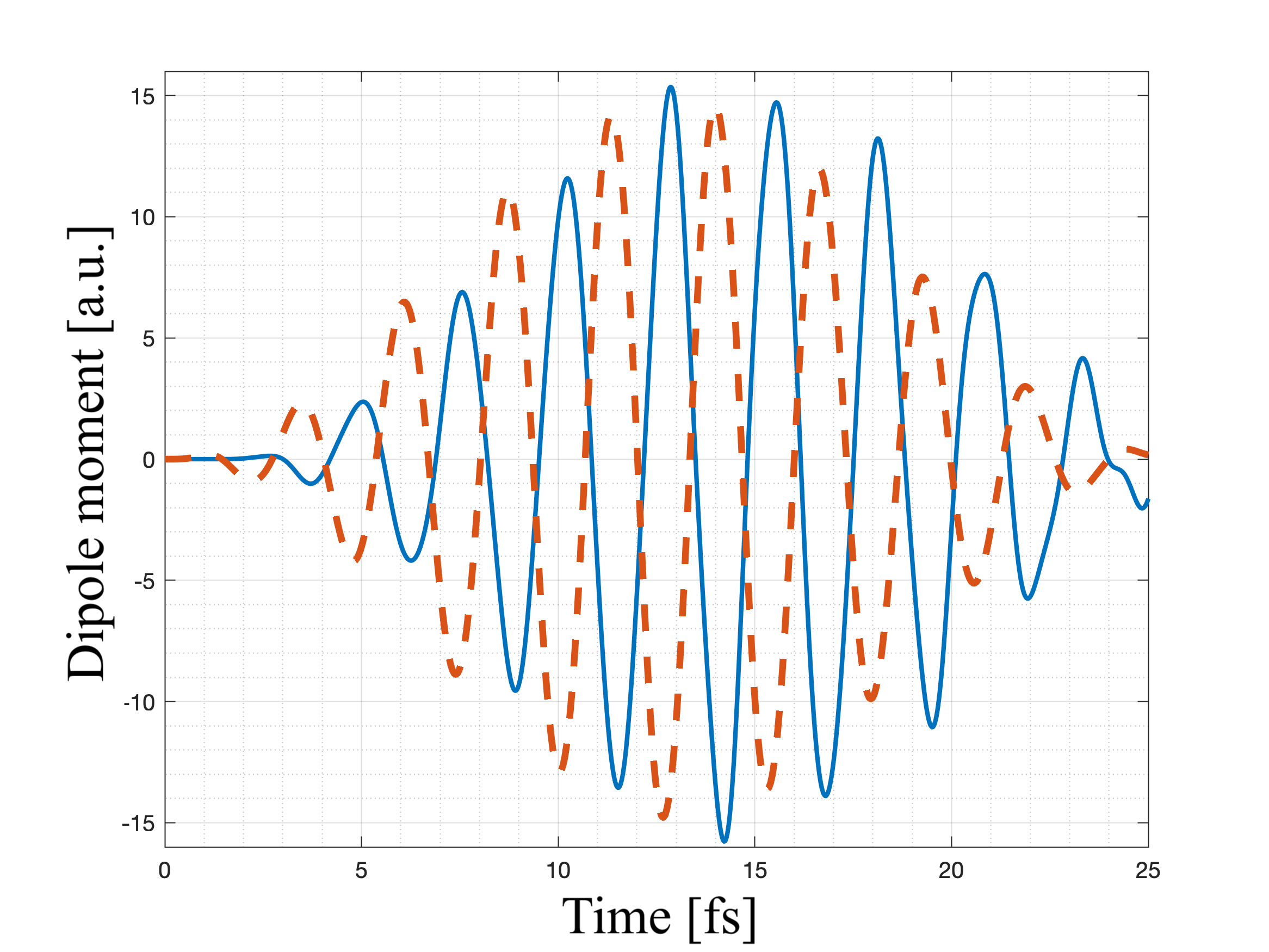}
\caption{
Comparison of the dipole moment along the C$_{12}$H$_{4}$ molecules axis ($\hat{z}$ direction), with separation of D=7500 a.u. ($\approx \lambda/2$ nm) between the domains. in red (dashed) the domain with the applied laser (Domain 1). Domain 2 (blue) is normalized with the factor of the dipole radiation for comparison.}
\label{fig:c12h4_7500_td}
\end{figure}

\begin{figure}[h]
\centering
\includegraphics[width=0.5\textwidth]{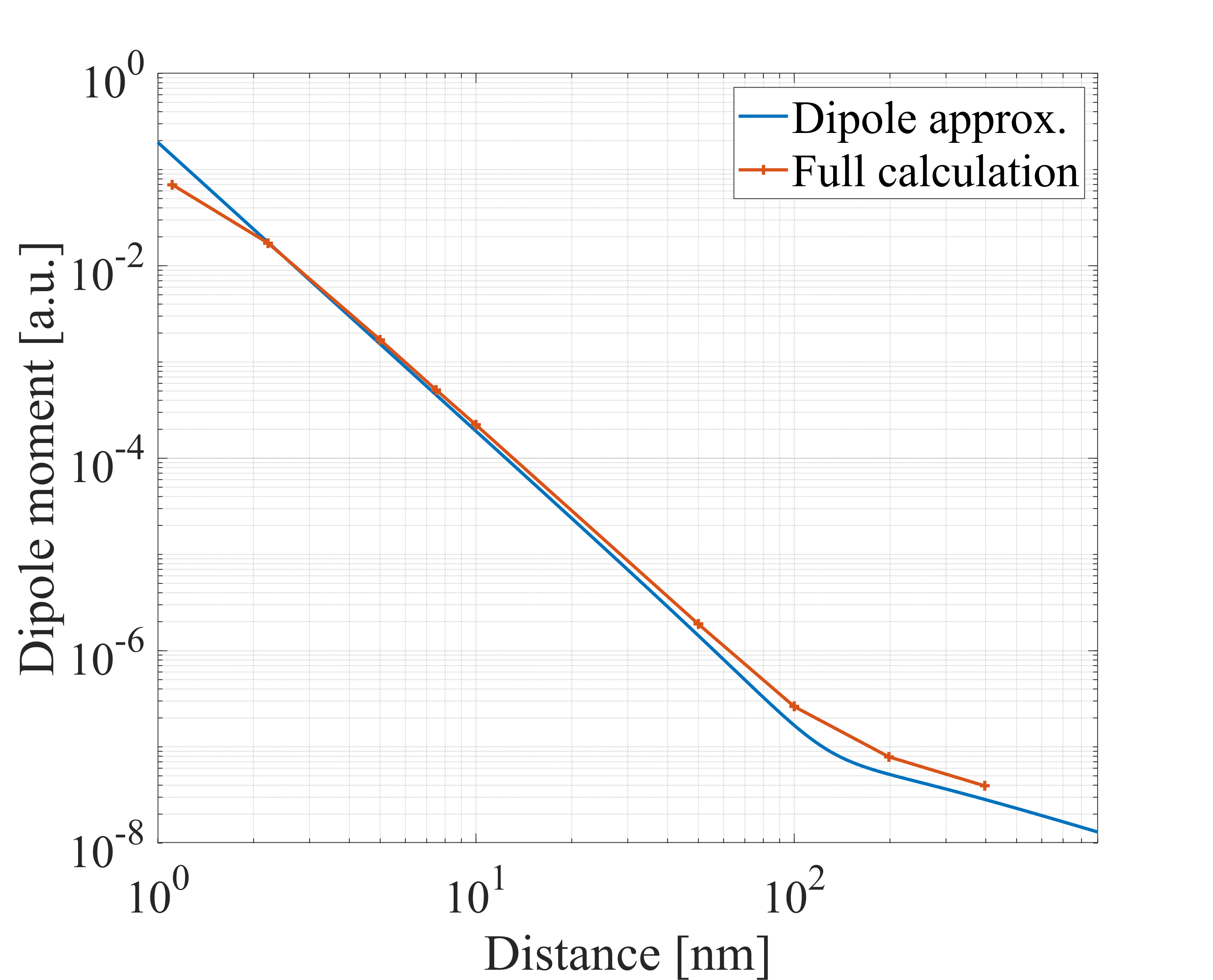}
\caption{The second domain induced dipole moment along the C$_{12}$H$_4$ molecules axis ($\hat{z}$ direction) versus the dimer separation,
the full calculation is shown in red, the result of the dipole approximation is shown in blue for comparison.}
\label{fig:c12h4_distances}
\end{figure}

The time evolution of both domain dipole moments is shown in Figs. \ref{fig:c12h4_90_td} and \ref{fig:c12h4_7500_td} at the chosen distances of $d=90$ a.u., $d=7500$ a.u. ($\approx {\lambda}/{2}$) respectively.
To scale both dipole moments, we normalized the amplitude of the domain without the laser by the factor of the dipole radiations envelope peaks  ${(\alpha_{z,z}{E}_{1,z}/D_{1,z})}^{-1}$ (Eq. \ref{polarizability as function of electric field}). The results match well with the amplitude and phase predicted by the dipole approximation of Eq. (\ref{eq: electric field inf dipole}).
A systematic analysis of the deviations from the dipole approximation is depicted in Fig. \ref{fig:c12h4_distances} for a range of distances from $20$ to $7500$ a.u. We observe several behaviors: for large distances, (above 90 a.u. in this example), we can see that the dipole radiation approximation accurately predicts the dipole moment response of domain 2. At shorter distances, where the domains are very close, we observed a significant deviation from the Hertzian dipole model because the molecules are too big relative to the separating distance. This is expected as in this regime the approximation is not valid. Finally, At very large distances there are also some small deviations between the full calculation and the approximation. Those deviations are caused by numerical errors as the signal becomes too weak.

\begin{figure}[h]
\centering
\includegraphics[width=0.5\textwidth]{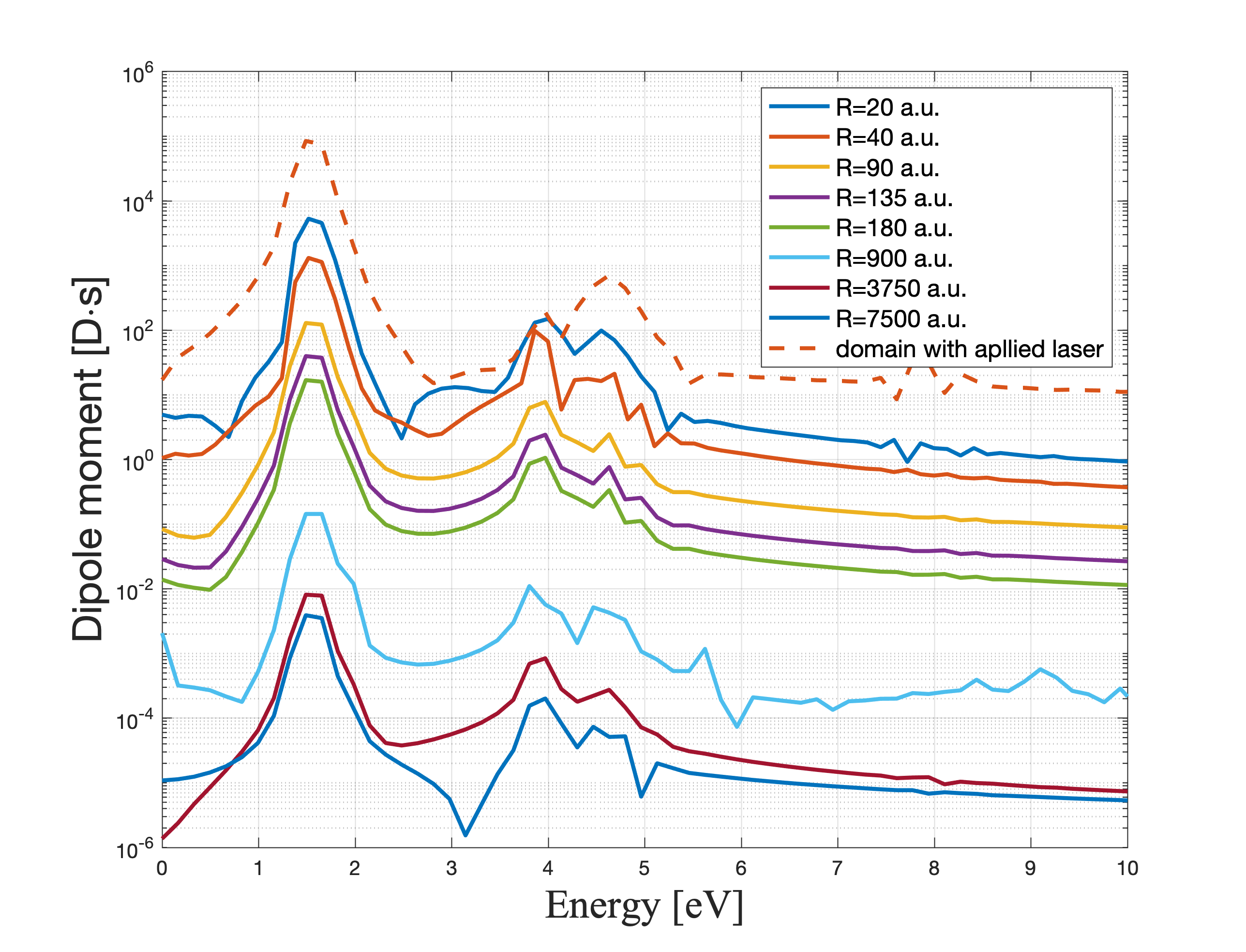}
\caption{Optical spectra of the C$_{12}$H$_4$ dipole moment for a sine squared pulse as a function of energy. Calculated for Domain 1 with the applied laser (in red - dashed) and for Domain 2 which is affected by the dipole-dipole radiation at various distances (colored).}
\label{fig:c12h4_fft}
\end{figure}

In Fig. \ref{fig:c12h4_fft}, the Fourier transform of the induced dipole moment is shown for Domain 1  (red dashed line) and for Domain 2 (colored lines) at several distances. 
The main peak is located as expected at $~1.54 $ eV which is the photon energy of the applied field. The dipole moment strength of Domain 2 is fading as the distance increases, as predicted by the dipole radiation approximation of Eqs. (\ref{eq: electric field inf dipole}) and (\ref{polarizability as function of electric field}). The second peak at $~4.0$ eV corresponds to a molecular resonance peak, while the third peak at $~4.65$ eV is the result of the non-linear effect due to the third harmonic of the electromagnetic pulse. This was verified by reducing the laser amplitude and observing a decrease by a cubic factor of the dipole moment peak at this peak energy.

\section{Summary}
We presented a multi-domain scheme for the calculation of the dynamics of a large system with separable densities. This scheme included the evaluation  of the electromagnetic retardation effects within TDDFT. In this method, the KS orbitals of each domain in the system are evolved in parallel in time  while the time dependent potentials are calculated by summing all sub-domains' retarded contributions. We showed a numerically efficient way to evaluate the integral expressions for the calculation of the retarded potentials using the Lorenz gauge. This approach was demonstrated in real space for two cumulene (C$_{12}$H$_4$) molecules, affecting one another. 
As predicted, for large distances, the calculated dipole moment of the far molecule, can be approximated by the response to a field of a Hertzian dipole radiation.
We analyzed the computational speedup, relative to a single domain, for the case of two nano-wires. The multi-domain scheme simulation time is not affected by the distance between the two molecules while the single-domain simulation time increases quasi quadratically with the distance. Another advantage of the method is its amenability to parallelization. Every domain can run on a different processor and requires only the implementation of an efficient data transfer mechanism.
An additional speedup can be achieved by using the multipole approximation for the retarded induced potential.

The developed method can be applied to several molecules. A computational speedup is achieved mostly when there is a large enough space between the molecules. A possible application can be an arrangement of molecules which  operate as a nano-antenna array.  

\appendix
\section{Lorenz gauge within TDDFT} \label{appendix:A}
TDCDFT maps the time dependent Schr\"odinger equation into effective single particle KS like equations in a gauge invariant way \cite{vignale2004mapping}. The dynamic Hamiltonian $\hat{\mathrm{H}}_{\mathrm{dyn}}$
resulting from the Kohn-Sham approach in TDCDFT can be written in the velocity gauge as \cite{escartin2015towards,vignale2004mapping,gabay2020lorenz} (in a.u):
\begin{subequations}
\begin{equation}\label{tdcdft}
    \hat{\mathrm{H}}_{\mathrm{dyn}}=\frac{1}{2}\bigg. \bigg(\frac{\nabla}{i} -\frac{q_e}{c}\textbf{A}_{\mathrm{s}}[\textbf{j}](\textbf{r},t)\bigg. \bigg)^2+v_{s}[{\rho}](\textbf{r},t)
\end{equation}
%+\nonumber \\
\begin{align}
&\textbf{A}_{\mathrm{s}}[\textbf{j}](\textbf{r},t)=\nonumber \\ &\textbf{A}_{\mathrm{ext}}(\textbf{r},t)+\textbf{A}_{\mathrm{ind}}[\textbf{j}](\textbf{r},t)+\textbf{A}_{\mathrm{xc}}[\textbf{j}](\textbf{r},t)
\end{align}
\begin{align}
&v_{s}[{\rho]}(\textbf{r},t)=\nonumber \\
&q_ev_{\mathrm{ext}}(\textbf{r},t)+q_ev_{\mathrm{ind}}[{\rho}](\textbf{r},t)+v_{\mathrm{xc}}[{\rho}](\textbf{r},t)
\end{align}
\end{subequations}
Here,  $q_e=-1$ since we use atomic units. The scalar potential ${v}_s$ is composed of the external (and ion) scalar potential contribution ${v}_\mathrm{ext}$, the exchange-correlation scalar potential ${v}_\mathrm{xc}$, and the induced scalar potential ${v}_\mathrm{ind}$. While
the vector potential $\textbf{A}_s$ is composed of the external vector potential contribution, $\textbf{A}_\mathrm{ext}$, the exchange-correlation vector potential, $\textbf{A}_\mathrm{xc}$ and the induced vector potential $\textbf{A}_\mathrm{ind}$. In this work, we adopt the approach described in \cite{gabay2020lorenz}. Under the assumption of weak magnetic fields, we neglect the effects of the exchange-correlation vector potential ($\textbf{A}_\mathrm{xc}$). The induced scalar and vector potentials are both functionals of the sources. The electron density and the current density can be expressed in terms of the KS single particle orbitals as
\begin{subequations}
\begin{equation} \label{charge density}
    {\rho}\left({\bf{r}}{,}{t}\right) =\mathop{\sum}\limits_{{n} = {1}}\limits^{{N}_{\bf{o}}}{{{\psi}}_{n}^{\ast}}\left({\bf{r}}{,}{t}\right){{\psi}}_{n}\left({\bf{r}}{,}{t}\right),
\end{equation}
\begin{align}\label{current density}
  &{\bf{j}}\left({\bf{r}},{t}\right) =\nonumber \\ &\mathop{\sum}\limits_{{n} = {1}}\limits^{{N}_{\mathrm{o}}}\frac{1}{2i}\left[{\psi}_{n}^{\ast}\left({\bf{r}},{t}\right){\nabla}{\psi}_{n}\left({\bf{r}},{t}\right){-}{\psi}_{n}\left({\bf{r}},{t}\right){\nabla}{\psi}_{n}^{\ast}\left({\bf{r}},{t}\right)\right] \nonumber \\  &+\frac{1}{c}{\rho}\left({\bf{r}},{t}\right){\bf{A}}_{s}\left({\bf{r}},{t}\right),
\end{align}
\end{subequations}
and they are independent of the gauge fixing of the electromagnetic potentials. Equation (\ref{current density}) and (\ref{tdcdft}) form a circular dependence between the current density and vector potential. In this work, we use the current density at the current time to evaluate the vector potential of the next time step. We use small time step and verify small increments in the current density to assure that this approximation is justified.
The induced potentials can be expressed in the Lorenz gauge to obtain the integral expression used in Eq. (\ref{4v potential}).

\section{Lorenz Gauge potentials}\label{appendix:B}
The induced four-vector potential in  integral expressions are given in Eq. (\ref{4v potential}). This section is provided for completeness.
The gauge-free fields can be expressed in term of the scalar, $v$, and vector, $\textbf{A}$, potentials as (atomic units):
\begin{subequations}
\begin{equation}\label{E}
\textbf{E}=-\nabla{v}-\frac{1}{c}\frac{\partial\textbf{A}}{\partial t}
\end{equation}
\begin{equation} \label{B}
\textbf{B}=\nabla \times \textbf{A}.
\end{equation}
\end{subequations}
The pros and cons of the Lorenz gauge fixing are described in \cite{gabay2020lorenz}. Here, we derive the resulting integral expressions \cite{balanis2012advanced}. The Lorenz gauge fixing condition is given by:
\begin{equation}\label{lorenz gauge fixing}
\nabla{\cdot \textbf{A}}+\frac{1}{c}\frac{\partial{v}}{\partial{t}} = 0
\end{equation}
Substituting Eq. (\ref{lorenz gauge fixing}) into the divergence Gauss's equation we obtain the scalar potential wave equation:
\begin{align}\label{scalar potential diff}
\nabla\cdot\textbf{E}&=-\nabla\cdot(\nabla{v}+\frac{1}{c}\frac{\partial\textbf{A}}{\partial t})\nonumber \\&= (\frac{1}{c^{2}}\frac{\partial^2}{\partial t^2}-\nabla^2)v=4\pi{q_e}\rho,
\end{align}
where $\rho$ is calculated as in (\ref{charge density}). The vector potential wave equation can be derived by substituting Eqs. (\ref{B}) and (\ref{E}) in Ampere's law:% by taking the curl of both sides of Eq. (\ref{B}):
\begin{align} \label{Amphere's law, gauge free}
&\nabla{\times{\textbf{B}}}=\nabla{\times \nabla{\times \textbf{A}}}= \nabla{\nabla{\cdot\textbf{A}}}-\nabla^2{\textbf{A}}=\nonumber \\ &{\frac{4\pi}{c}}q_e\textbf{j}+\frac{1}{c} \frac{\partial\textbf{E}}{\partial t}={\frac{4\pi}{c}}q_e\textbf{j}+\frac{1}{c} \frac{\partial}{\partial t}({-\nabla{v}-\frac{1}{c}\frac{\partial\textbf{A}}{\partial t}}),
\end{align}
where $\textbf{j}$ is calculated as in (\ref{current density}). Further rearranging Eq. (\ref{Amphere's law, gauge free}) and applying the Lorenz gauge fixing condition (\ref{lorenz gauge fixing}), we obtain:
\begin{align}\label{vector potential diff} 
(\frac{1}{c^2}\frac{\partial^2}{\partial t^2}-\nabla^2)\textbf{A}&={\frac{4\pi}{c}}q_e\textbf{j}-\nabla(\frac{1}{c}\frac{\partial}{\partial t}{v}+\nabla\cdot\textbf{A})\nonumber \\&={\frac{4\pi}{c}}q_e\textbf{j}.
\end{align}
Equations (\ref{vector potential diff}) and ( \ref{scalar potential diff}) are the inhomogeneous electromagnetic wave equations, whose solutions are known as the retarded Lorenz gauge potentials. The integral forms of the solution of (\ref{vector potential diff}) and (\ref{scalar potential diff}) are given by:
\begin{equation}\label{general scalar potential}
   {v}(\textbf{r},t)=
   q_e\int{\frac{\delta(t'+\frac{R}{c}-t)}{R}}{\rho(\textbf{r}',t'})d\textbf{r}'d{t'}
\end{equation}
\begin{equation}\label{general vector potential}
  \textbf{A}(\textbf{r},t)=
   q_e\int{\frac{\delta(t'+\frac{R}{c}-t)}{R}}\frac{\textbf{j}(\textbf{r}',t')}{c}d\textbf{r}'d{t'}
\end{equation}
The results of (\ref{general scalar potential}) and (\ref{general vector potential}) describe the solution for the induced potentials which are used in Equations (\ref{4v potential}) and in appendix \ref{appendix:A}. The external potentials fulfill the homogeneous form of Equations (\ref{scalar potential diff}) and (\ref{vector potential diff}) which do not include the charge and current densities.  \section{Nitrogen}\label{appendix:c}
\begin{figure}[h]
\centering
\includegraphics[width=0.4\textwidth]{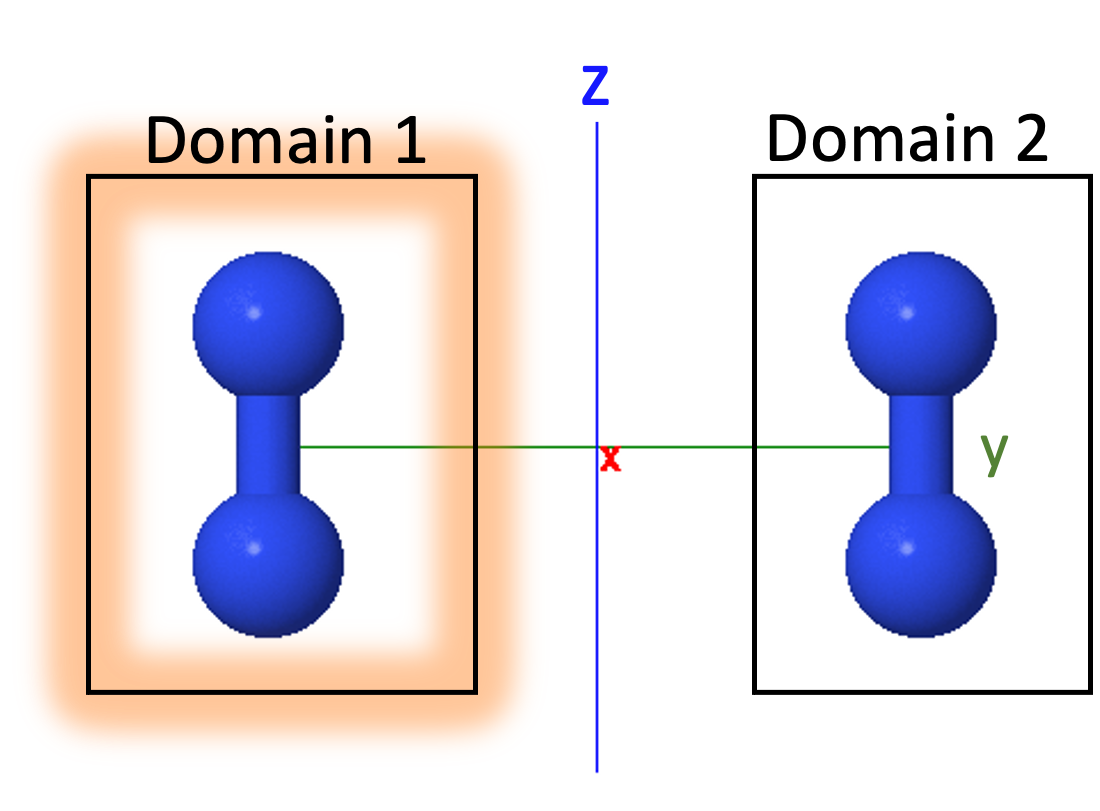}
\caption{An illustration of the nitrogen multi-domain setup. Two sub-domains of N$_{2}$ set in parallel to their axes, located at various distances. An external field is applied to Domain 1. Domain 2 is affected by the induced potential of Domain 1.}
\label{fig:dimers N2}
\end{figure}

We also calculated the response of a nitrogen gas N$\equiv$N, see Fig. \ref{fig:dimers N2}.
The N-N bond length is 1.037 a.u. 
In the following, the domain size of each sub-domain is a box of 20$\times$20$\times$20  a.u. with a grid spacing of $0.4$ a.u.
We used absorbing boundary conditions (ABC) with a boundary layer of $5$ a.u. on each spatial dimension and exponential damping coefficient of $0.25$ a.u. The electronic structures were evolved in time for $30$ femtoseconds with a time step of $1$ attosecond. These parameters were all specified to ensure numerical correctness of the model.

\begin{figure}
\centering
\includegraphics[width=0.5\textwidth]{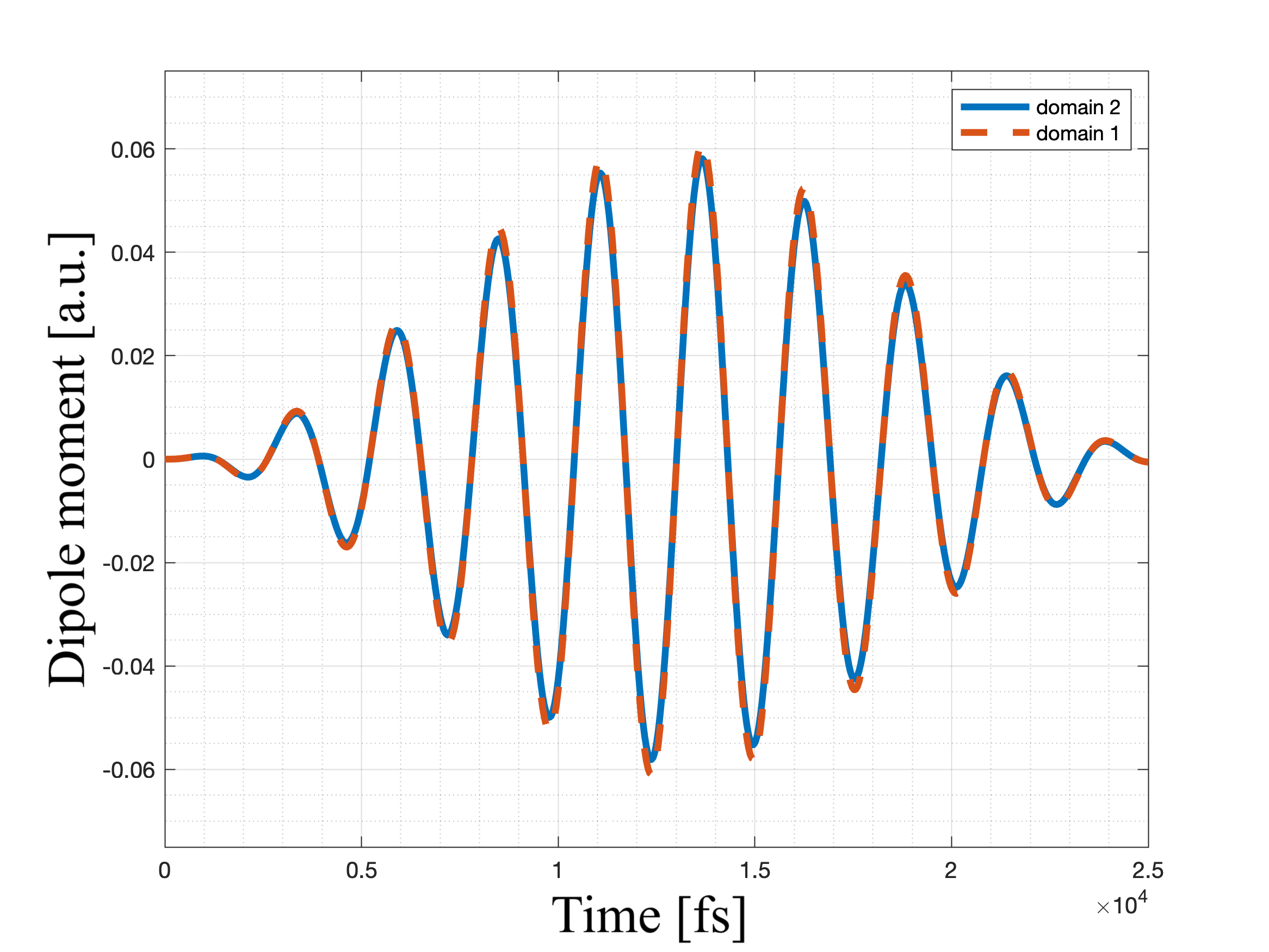}
\caption{Comparison of the dipole moment along the N$_{2}$ molecules axis ($\bf{\hat{z}}$ direction), with separation of $d=22$ a.u. between the domains. in red (dashed) the domain with the applied laser. The second domain (blue) is normalized with the factor of the dipole response for comparison.}
\label{fig:N2_22_td}
\end{figure}

\begin{figure}
\centering
\includegraphics[width=0.5\textwidth]{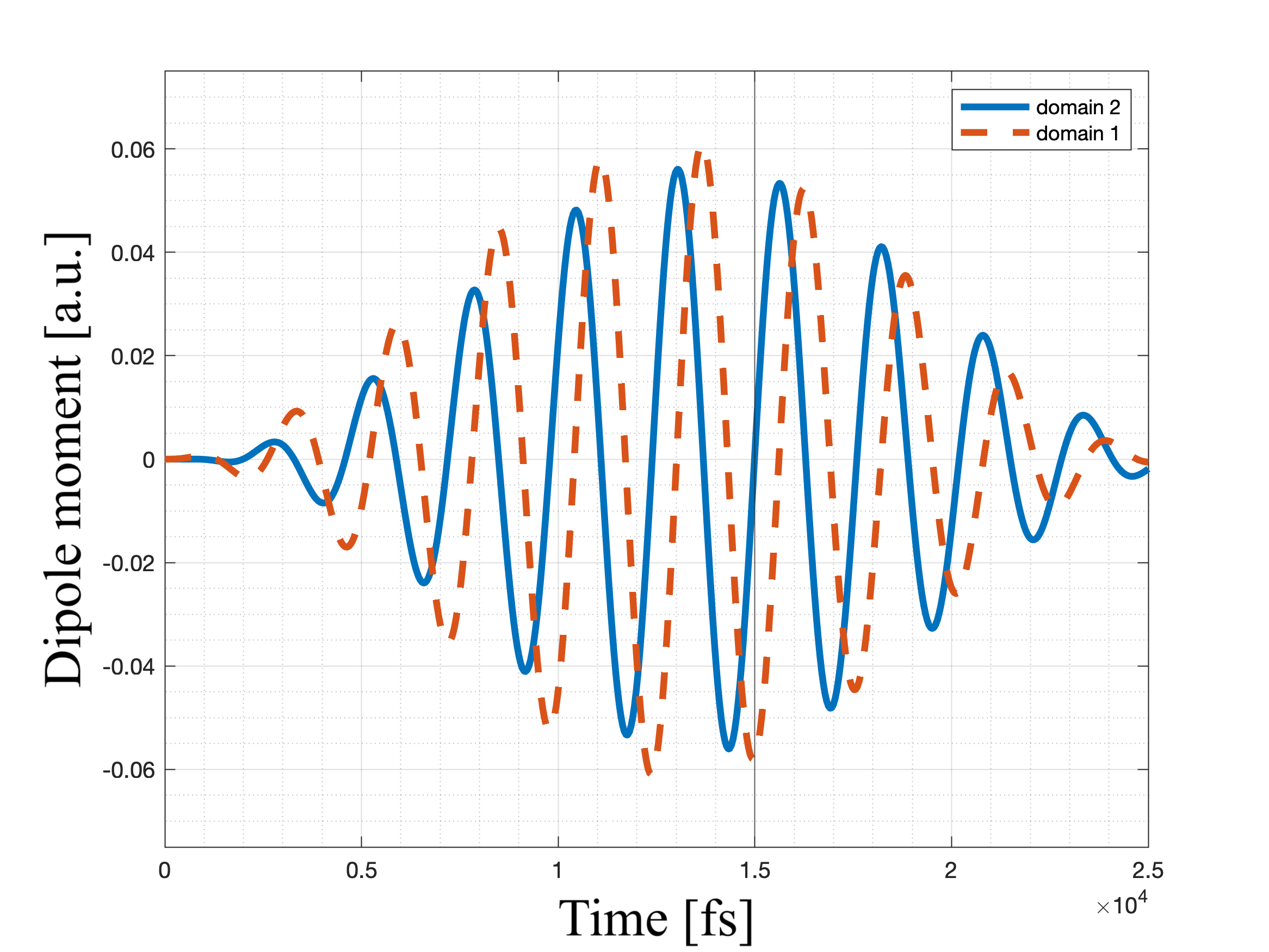}
\caption{Comparison of the dipole moment along the N$_2$ molecules axis ($\bf{\hat{z}}$ direction), with separation of d=3700 a.u. between the domains. In red (dashed)- the domain with the applied laser. The second sub-domain (blue) is normalized with the factor of the dipole response for comparison.}
\label{fig:N2_3700_td}
\end{figure}

The time evolution of
both domains' dipole moments is shown in Figs. \ref{fig:N2_22_td}, \ref{fig:N2_3700_td} at the chosen distances of $d=22$ a.u., and $d=3700$ a.u. ($\approx{\lambda}/{4}$), respectively.
To scale both dipole moments, we normalized the amplitude of the domain without the laser by the factor of the dipole radiations ${(\alpha_{z,z}{E}_{1,z}/D_{1,z})}^{-1}$ (Eq. (\ref{polarizability as function of electric field})). The results match well with the amplitude and phase predicted by the dipole approximation of Eq. (\ref{eq: electric field inf dipole}).

\section*{Acknowledgement}
Financial support by the IEEE Antennas and Propagation Society Graduate Fellowship Program - Quantum Technologies Initiative is gratefully acknowledged.
This research was also supported by Grant No. 2018182 from the United States--Israel Binational Science Foundation (BSF).

\providecommand{\noopsort}[1]{}\providecommand{\singleletter}[1]{#1}%

\end{document}